\documentclass{icrc29}
\usepackage{graphicx,amssymb,amsmath,times}

\begin{document}
\title[Production mechanisms \ldots]{Production  mechanisms of multiple primaries for Cosmic Rays Showers}

\author[G. Imponente and G. Sartorelli] {Giovanni Imponente $^a$ and
             Gabriella Sartorelli $^b$ 
	 \\
        (a) Museo Storico della Fisica, Centro Studi e Ricerche ``E.Fermi'' 
        and \\ Dipartimento di Fisica Universit\`a ``La Sapienza'', 
        P.za A. Moro 2 - 00185 Roma, Italy \\ 
        (b) Dipartimento di Fisica Universit\`a di Bologna and INFN, 
        Sez. di Bologna, Italy
        }
\presenter{Presenter: G. Imponente (gpi@physics.org), \  
uki-imponente-G-abs2-he23-poster}


\maketitle

\begin{abstract}
We investigate the physical mechanism of the GZ-effect 
that could explain  the 
production of multiple
 primaries from an event initiated outside the Earth's 
 atmosphere. In  this case,  there  would correspondingly be 
 multiple extensive air showers in temporal
 coincidence at ground, even for detectors separated by
 many kilometers, and also showers initiated by primaries 
 of different  energies could  consequently  have a common source.
 We analyse the perspectives and limits of some models and 
 discuss the  experimental  counterparts.
\end{abstract}

\section{Introduction}

The history of a Cosmic ray from the production point 
and through the acceleration sites undergoes many 
changes in velocity and eventually in chemical structure. 
The nature of such a primary, either a heavy nucleus or proton
or gamma-ray or whatever, influences strongly the byproduct of
its interaction with the medium: interstellar matter, 
Cosmic
Microwave Background photons (CMB)  or local 
interstellar/intergalactic/galactic magne\-tic fields 
\cite{GZK66,PSB76,MS98,SS99,BILS02}.
A  great interest is therefore devoted to understanding  the abundances
of protons and relative chemical composition of the CR's flux, 
in terms of other elements or ions. \\
Among the various kinds of projectiles hitting the Earth's 
atmosphere, we will consider the fragments deriving from the 
photo disintegration of heavy nuclei (for example Fe) 
when interacting with 
the solar magnetic field \cite{GZ60}: the crucial aspect of this 
fragmentation relies in the possibility of detecting on Earth
two (or eventually more) of them. 
In fact, the influence of the solar 
field permeates the space surrounding the Sun up 
to distances limited to 3 or 4 AU. This relatively 
small volume allows the fragments to arrive on Earth 
almost simultaneously and spaced ranging some  up to 
few thousands Km. 
Thus, the Extensive Air Showers (EAS)
 generated by the two projectiles when hitting the atmosphere
would be temporally as well spatially correlated and detectable
when having many  detectors placed at 
different distances, some closely and some widely 
spaced. \\
The arrival rate computed in the present paper
heavily depends on the 
energy of the incoming particles and for this reason 
we include in our estimate small variations (some units)
providing important differences in the expectation 
values. \\
What is strongly encouraging in the experimental search 
for this peculiar phenomena is the possibility of 
being detected by 
the new experiment ``Extreme Energy Events'' (EEE)
which is starting in Italy \cite{eee}. 
In fact, the disposition of the 
particle detectors is planned  inside numerous 
High Schools over all the Italian territory (about 
300.000 Km$^2$), more densely inside the cities, 
from south to north.

\section{Heavy primaries and correlation}

\subsection{Spatial and temporal correlation}

The study and the detection of
(ultra-)high energy CRs in temporal 
coincidence  is based on the energy of the primaries, 
the time difference 
 between the two (or more) events, the reconstructed 
 arrival directions, the possible source or set of sources.
Each of these patterns is strongly related to the others, 
and the accuracy in reconstructing a specific air 
shower  puts forward experimental results paired with 
theoretical shortcomings.\\
Gerasimova and Zatsepin \cite{GZ60}, in the early 
Sixties,  proposed a theoretical prediction on large 
scale correlations: the so-called
GZ effect describes the disintegration of CR nuclei  
in the field of solar photons, leading to the formation of extended 
time-correlated EAS pairs with core distances (their estimate)
 $\sim 1 $ Km.\\
 Medina-Tanco and Watson \cite{MTW99}
have re-evaluated the background of solar photons, 
 in view of a simple model for the solar radiation field
and of the existing/planned (at that time) 
experiments (like OWL, AGASA and AUGER).
The estimates were  limited by  the 
blindness of the detectors during the Earth's 
day side exposition, 
which is the condition for higher flux (the former), 
the smallness of the array (the second) 
or the small acceptance for showers at large distances
for the latter. 

\subsection{Heavy nuclei and origin of fragments}

Basically, 
multiple correlated showers, temporally as well spatially, 
can  arise  either from  local astrophysical events, 
on the distance scale of
the solar system, or from exotic phenomena, 
$E> 10^{15}$ eV air showers, which  may require anomalous 
processes high in the atmosphere.\\
The electromagnetic size $N_e$ and the muon number $N_{\mu}$ of a shower, 
depend differently on the energy and mass of the primary:
these allow an estimate of $E$ and $A$ for each shower. 
Qualitatively, since 
%
$N_{\mu} \sim K^{\prime} A^{1 - \beta/\alpha} N_e^{\beta/\alpha} $
with  $\alpha >1$, $\beta <1$, heavy primaries can 
be selected choosing muon rich showers. 
Evaluating $N_e$, a proton shower can be associated to a small 
muon number, while the highest muon multiplicities are related
to iron nuclei, therefore the measurement of $A$ for the 
single EAS would be the definitive signature for 
the observation of a fragmentation process event.
The  photo disintegration \cite{PSB76}
of  heavy nuclei 
with a photon background are dominantly 
in the channels
$A + \gamma_{} \rightarrow (A-1) + N$, or 
								$\rightarrow (A-2) + 2N $ 
being $N$ a nucleon, and most of the absorption cross section 
results in the emission of only  single
nucleons, either protons or neutrons, while 
pair production 
processes  will not give rise to multiple primaries.
%
%
The fragmentation process goes through the 
photo nuclear interaction with an individual nucleon, 
and refers to measurements and values of the 
nuclear-collision cross sections
at energies lower than CRs', 
so that 
could need corrections when applied  to CRs
photo disintegration. \\
The  energy change is related to that of the atomic number
as ${\Delta E}/{E} = {\Delta A}/{A}$
hence, for example, a Fe nucleus losing a proton diminishes
its energy of less than 2\%. 
The most evident effect for single or multiple 
disintegrations off photon backgrounds (of different kinds)
is a wide richer variety in chemical composition 
of CRs flux, together with the protons' one.

\subsection{Current searches and methods}

The look for correlations between CRs aims 
to distinguish some common
features, such as the sources or their number and distribution, 
and the characteristics of the propagation to the Earth. 
The first unusual simultaneous increase 
in the CRs shower rate 
 has been reported in 1983 \cite{FMO83}; 
 an experiment devoted to the search 
for correlations between primaries
has been proposed by Carrel and Martin in 1994 \cite{CM94}
who wanted (with no success) to measure showers arising 
from multiple primaries originated by a hypothetical 
single (or multiple) event far from the Earth.\\
Many experiments are running looking for UHECRs, but to date there
are no stringent data about spatial or temporal clustering, 
apart from few signals from AGASA \cite{Taetal99}  
and low statistics for anisotropies \cite{ST95}. 
%
%
%
%
%
%
A recent search for time correlated events is given by 
the CHICOS project \cite{chicos}, 
 currently  running 
in California High Schools, covering an area of about 400 Km$^2$,
providing one observed candidate event, yet compatible 
with an accidental coincidence 
of independent showers. 
The Large Area Air Shower (LAAS) \cite{Oetal01_1} in Japan 
has  a  shower array of about 130.000 Km$^2$ area;
the  group reported 
one pair of EAS
with a very small time and angular  difference. 
They found  
\cite{Oetal03_1},
four coincident
event candidates,  though with very low significance.

\section{Gerasimova-Zatsepin effect}

The photodisintegration process can happen at various distances
from Earth, ranging about 0.04 AU up to 4 AU off the solar magnetic 
field \cite{MTW99}. 
Given a nucleus of mass $M$ 
and mass number $A$, the Lorentz factor is 
$\gamma = E/(A m_p c^2)$. 
Then, an 
isotropic emission of nucleon(s) in the reference system of 
the parent nucleus would result
in a cone with aperture of $\sim 1/\gamma$ around the 
original direction of propagation, when viewed 
in the reference system 
of the Earth. 
For high values of the Lorentz factor ($\gamma > 10^7$), 
both fragments have exactly the same direction as the 
incoming nucleus, after the 
interaction with the photon:  
the fragments will proceed on their way almost
parallel and will be deflected under the action of the  
magnetic field, depending on their charge,
producing a core separation  expected at Earth 
that can be of the order of many kilometers. 
 The further the disintegration takes place, 
the larger will be the distance between the fragments (and of the 
produced showers) when entering the atmosphere.\\
We re-evaluate the rate of secondary CRs, allowing a
small energy variation of the parent primary and scanning all
directions in the sky, night-side and day-side 
(much more favorable), in view of
ongoing and new experiments, such as EEE \cite{eee}, which, 
once in operation, can allow a consistent source 
of investigation for simultaneous showers over a wide range
of distances, up to hundreds of  kilometers
away. \\
Given the mean free path of a nucleus against 
photo disintegration, the ratio 
$\eta_{GZ} = {\Phi_{GZ}}/{\Phi_{\infty}}$
between the flux $\Phi_{GZ}$
of the GZ fragment pairs
 and  the unperturbed
incoming one $\Phi_{\infty}$ 
is related to 
the parameter $\delta$, the average 
separation between the showers' cores when arriving on the Earth, 
spanning from about 1 to 2000 Km. 
Small values of 
$\delta$ refer to events originating in the vicinity 
of the Earth, larger values refer to further distances. \\
The EEE project \cite{eee} will  cover a surface 
 $A_{EEE} \cong 3 \cdot 10^5$ Km$^2$.
In principle, since the GZ effect is for 
well spaced showers (of the order of Km or tens of),
 the whole array 
does not need to be  very dense.\\
Let us  consider a primary Fe nucleus with energy 
varying between $E\sim 10^{17}$ and $E\sim 10^{19}$ eV. 
%
%
\begin{table}[htbp]   
\begin{center}
\begin{tabular}{ccc}  
\parbox[t]{75mm}{
		\begin{tabular}{||c|cc|cc||} 
		\hline \hline

 			 $\delta$ (Km)    &  $\eta_{GZ} $ &&  event rate &  \\
 	                            \hline \hline
 	                            && $\times10^{-5} $&&\\
			 { 2 $\div$ 5 }   & 0.08           &            &    66  - 112 &\\
                               
				5 $\div$  6       &  0.40        &  &     330 - 562 &\\
				$>$ 6					& 0.4 &			&						33 - 56 &\\
                             \hline
                             \hline
                      && & & $\times 10^3$\\
				$<$ 250        &  1.6     & &  1.3 - 2.2 &  \\
										
			250 $\div$ 400      & 2.5       & &   2.1 - 3.5  &\\
											\hline
			400 $\div$ 600  	 	&	5.0	&	&		 4.2 - 7.1   &\\
			600 $\div$ 1000  		& 13 	&	& 	10.5 - 17.8  &	\\
			\hline
			1000 $\div$ 1500 		& 32  &	& 	26.2 - 44.6  	&	\\
				$>$ 1500            & 50 		&& 	 41.6 - 70.7  	&	\\
                             \hline
		\end{tabular}
		\caption{\label{gzflux1} 
		GZ events rate for various $\delta$, 
		with primary's energy $3.2\div 7.9 \cdot 10^{17}$ eV, 
		integrated over $A_{EEE}$, in one year.}
} &
& \parbox[t]{70mm}{
%
\begin{tabular}{||c|cc|cc||} \hline \hline

$\delta$ (Km) &     $\eta_{GZ}   $  &&  event rate &  \\
                             \hline \hline
                             && $\times 10^{-8}$&&	$\times 10^{-2}$	\\
$\gg 50$          &  50.1   &      &   0.8 - 19.5 & \\
                             \hline
$>$ 50        &  20.0       & &     0.3 - 7.8 &\\
                             \hline
30 $\div$ 50        &  7.9 						& 	& 0.1 - 3.1	&	\\
 \hline
20 $\div$ 30        & 3.2			& & 0.05 - 1.2	&\\
\hline
 10 $\div$ 20       &  5.0    &&  0.08 - 1.9 &\\
 \hline
 5 $\div$ 10  				& 2.0 	&	&	0.03 - 0.7 &\\
 \hline
 $<$ 5 				& 1.3			& 	&	0.02 - 0.5 &\\
                             \hline
\end{tabular}
\caption{\label{gzflux4} 
GZ events rate for various  $\delta$, 
given a primary's energy ranging 
$1.0\div 6.3 \cdot 10^{19}$ eV, integrated over
$A_{EEE}$, in one year.}
	}
	\end{tabular}
%
\end{center}
\end{table}
On the basis of the fraction 
$\eta_{GZ}$ of the total CRs flux,
 we  compute the integral flux, over $A_{EEE}$ and for one year. 
One can look at different  separations between the fragments at 
Earth, taking in mind that the spacing depends slightly also 
from the incoming direction in the sky.
The range given for the event rate is due 
to allowing $\eta_{GZ}$
being the same for slight variations of the primary's energy.
For example, in  Table \ref{gzflux1} we consider the two values of $E$, 
$3.2	$ and $ 7.9 \cdot 10^{17} $ eV: 
correspondingly,  we find the event rates for different ranges of 
$\delta$.	
Similarly, for the interval of energy $1.0 - 6.3 \cdot 10^{19}$ eV, 
and $2.5-3.9 \cdot 10^{18}$ eV we have the 
rates reported in Table \ref{gzflux4} and Table \ref{gzflux2}, respectively.\\
\begin{table}[h]   
\begin{center}
\begin{tabular}{||c|cc|c||} \hline \hline

		$\delta$ (Km) &     $\eta_{GZ}$ &  &  event rate   \\
                             \hline \hline
                             &&$\times 10^{-6}  $&\\
		20 $\div$ 60           &  1.58         &&   4 - 10 \\
                            \hline
	60 $\div$ 80        &  2.51        &&    6 - 17 \\
                             \hline
		60 $\div$ 80         &  6.31 					&	& 	 15 - 42 		\\
 		\hline
		80 $\div$ 100       &  15.8 					&	& 	 38 - 104 		\\
                             \hline
	\end{tabular}
		\caption{			\label{gzflux2} 			GZ events rate for various $\delta$, 
		with primary's energy $2.5\div 3.9 \cdot 10^{18}$ eV,
		integrated over $A_{EEE}$, in one year.
		Note that the  two rows with equal $\delta$ refer to different directions in the sky.}
\end{center}
\end{table}
\section{Discussion and conclusions}

The event rates are strongly dependent on the primary energy, 
as can be seen in the case of  a very energetic one, 
since the flux varies accordingly by orders of magnitude.
The fragments can generate showers over a wide range distances
between them, 
from few to thousands Km, thus not requiring a 
peculiarly spaced array of detectors, but moreover a 
large area. 
If we consider primaries with a given $E$ and flux, 
our calculations split essentially  in two classes: 
the first relies on the events characterized  by a large separation 
of fragments on Earth $\delta >200 $ Km, 
i.e. events originated closer to 
the solar influence, at distances of order $>2 ~AU$ from the Earth;
the second is given by a smaller separation of the fragments 
$\delta  < 15 $ Km, corresponding to primaries 
photo disintegrated at distances $< 2 ~AU$.\\
Considering a relatively modest variation for the incoming particle 
energy (about few units) 
and for the values computed for $\eta_{GZ}$, we find 
that the rate of GZ events expected can be effectively 
high (see Tables), leaving very promising work for the observation
of the GZ effect. 		
			%



\begin{thebibliography}{99}


\bibitem{GZK66}
K. Greisen \textit{Phys. Rev. Lett.} \textbf{16}, 748 (1966).
G.T. Zatsepin and V.A. Kuzmin, \textit{JETP Lett.} \textbf{4}, 78 (1966).

\bibitem{PSB76}
J.L.Puget, F.W. Stecker and J.H.Bredekamp, 
\textit{Astrophys. Journ.} \textbf{205}, 638  (1976)


\bibitem{MS98}
M.A. Malkan and F.W. Stecker, 
\textit{Astrophys. Journ.} \textbf{496}, 13  (1998)


\bibitem{SS99}
F.W. Stecker and M.H. Salamon, 
\textit{Astrophys. Journ.} \textbf{512}, 521-526  (1999)


\bibitem{BILS02}
G. Bertone, C. Isola, M. Lemoine and G. Sigl, 
\textit{Phys. Rev. D} \textbf{66} 103003 (2002)


\bibitem{GZ60}
N.M. Gerasimova and G.T. Zatsepin, \textit{Sov. Phys. JETP}
 \textbf{11}, 899 (1960) 

\bibitem{eee}
``The EEE Project'',  these Proceedings.


\bibitem{MTW99}
G.A. Medina-Tanco and A.A. Watson, 
\textit{Astropart. Phys.} \textbf{10}, 157  (1999)




\bibitem{FMO83}
D.J.Fegan, B. McBreen and C.O'Sullivan,
\textit{Phys.Rev.Lett} \textbf{51}, 2341 (1983)

\bibitem{CM94}
O. Carrel and M.Martin, \textit{Phys. Lett. B} \textbf{325}, 526 (1994)

\bibitem{Taetal99}
M. Takeda et al., \textit{Astrophys. Journ.} \textbf{522}, 225-237 (1999)


\bibitem{ST95}
T. Stanev et al.,
\textit{Phys. Rev. Lett.} \textbf{75}, 3056  (1995)








\bibitem{chicos}
B.E. Carlson et al.,
\textit{J.Phys. G} \textbf{31} 409-416 (2005), available 
astro-ph/0411212   (2004)

\bibitem{Oetal01_1}
N. Ochi et al.,
\textit{Nucl.Phys.B Proc. Suppl.} \textbf{97}, 165-168  (2001) ,
\textit{ibid.} 173-176





\bibitem{Oetal03_1}
N. Ochi et al.,
\textit{Nucl.Phys.B Proc. Suppl.} \textbf{122}, 333-336  (2003), 
\textit{ibid.}  337-340




\end{thebibliography}
\end{document}